\documentclass[aps,prl,reprint,twocolumn,floatfix,nofootinbib]{revtex4-2}
\renewcommand{\selectlanguage}[1]{}
\usepackage{amsmath,amssymb}
\usepackage{braket}
\usepackage{graphicx}
\usepackage{bm}
\usepackage{xcolor}
\usepackage{tikz}
\usetikzlibrary{circuits.ee.IEC,arrows.meta,decorations.pathmorphing,positioning}
\usepackage[colorlinks=true,linkcolor=blue,citecolor=blue,urlcolor=blue]{hyperref}
\graphicspath{{figures/}}

\newcommand{\Ec}{E_{C}}
\newcommand{\EJ}{E_{J}}
\newcommand{\EL}{E_{L}}
\newcommand{\phiext}{\varphi_{\mathrm{ext}}}
\newcommand{\Hflux}{\mathcal{H}_{\mathrm{F}}}
\newcommand{\Hmov}{\mathcal{H}_{\mathrm{mov}}}
\newcommand{\calH}{\mathcal{H}}
\newcommand{\Hquad}{\mathcal{H}_{\mathrm{quad}}}
\newcommand{\xqp}{x_{\mathrm{qp}}}
\newcommand{\OC}{\hat{O}_{c}}
\newcommand{\OS}{\hat{O}_{s}}
\newcommand{\phiJ}{\hat{\varphi}_{J}}

\newcommand{\qpratefix}[1]{{\color{blue}#1}}
\newcommand{\newroundedit}[1]{{\color{blue}#1}}

\begin{document}
	
	\title{Floquet Quasiparticle Poisoning of Frozonium}
	\author{Haoyu Guo}
	\author{Debanjan Chowdhury}\email{debanjanchowdhury@cornell.edu}
	\affiliation{Department of Physics, Cornell University, Ithaca, NY 14853, USA}
	
	\date{\today}
	
	\begin{abstract}
		Periodic driving can suppress the Josephson nonlinearity of a fluxonium superconducting circuit, producing a nearly harmonic Floquet spectrum at isolated freezing points \href{https://www.cell.com/newton/fulltext/S2950-6360(26)00036-8}{[K. Lewellen et al., Newton {\bf 2}, 100434 (2026)]}.
		Here we show that this dynamically frozen behavior does not generically suppress quasiparticle-induced dissipation in the resulting frozonium circuit. 
		We formulate quasiparticle processes in the frozonium using a Floquet framework and analyze both drive-assisted Cooper-pair breaking and tunneling of pre-existing quasiparticles. 
		Pair generation is controlled by gap-breaking thresholds at high drive frequencies, while multiphoton resonances produce pronounced rate enhancements at lower frequencies. 
		Quasiparticle tunneling exhibits connected resonance structures organized by the harmonic Floquet-Magnus spectrum near the freezing point, with resonant hybridization generating characteristic avoided crossings. 
		Our results show that suitable operating regimes must balance dynamical freezing against quasiparticle loss and provide a framework for identifying experimental drive parameters away from harmful resonances.
	\end{abstract}
	
	\maketitle
	
	\textit{Introduction.---} Inductively shunted Josephson circuits in the fluxonium regime~\cite{Manucharyan2009Fluxonium,Koch2009} have emerged as a leading platform for high-coherence superconducting qubits~\cite{Nguyen2019HighCoherence,Somoroff2023Millisecond,Ding2023FTF}, with millisecond relaxation times now demonstrated in 3D geometries~\cite{Somoroff2023Millisecond} and high-fidelity two-qubit gates demonstrated
	in planar geometries~\cite{Ding2023FTF}. 
	Their large superinductance suppresses sensitivity to offset charge noise while preserving strong anharmonicity, giving access to qubit frequencies well below those of conventional transmons~\cite{Masluk2012Superinductance,Koch2007,Blais2021, ficheux_fast_2021}. 
	Periodic driving offers an additional, dynamically tunable axis of control that can be used to access new regimes of the fluxonium circuit. 
	For instance, microwave modulation can engineer Floquet dynamical sweet spots that suppress dephasing from low-frequency noise~\cite{Didier2019,Mundada2020FloquetFluxonium,Huang2021DynamicalSweetSpots}, and similar Floquet ideas have been used to protect interactions in coupled fluxonia and related circuits~\cite{Thibodeau2024FloquetMolecule,Wang2024}.

	The recently introduced \emph{frozonium} regime~\cite{Lewellen2025Frozonium} sharpens this control principle and connects it to the broader phenomenon of dynamical freezing~\cite{Das2010Freezing,Haldar2021Freezing, HGuo2025, RMukherjee2026,Lu2026}. At isolated freezing points of the drive amplitude, the high-frequency Floquet-Magnus expansion shows that the leading Josephson contribution to the effective Hamiltonian is suppressed parametrically in the inverse drive frequency, so that the low-energy spectrum of the driven fluxonium reduces to that of a linear bosonic oscillator~\cite{Lewellen2025Frozonium,Mukherjee2025,Eckardt2015,Bukov2015}.
	This dynamical control simultaneously suppresses chaos in coupled circuits~\cite{Mukherjee2025} and renders the level structure largely insensitive to slow flux noise~\cite{Anton2013FluxNoise}, with the parametric flux dispersion at the freezing point exponentially small in $\sqrt{8\Ec/\EL}$~\cite{Lewellen2025Frozonium}.
	However, the theoretical treatment of frozonium so far has focused on the {\it bosonic} Hilbert space, not on the Bogoliubov quasiparticle (QP) degrees of freedom associated with the superconductor that makes up the circuit.
	Previously, error mechanisms in superconducting circuits arising from quasiparticles that tunnel across the small junction have been analyzed extensively~\cite{Catelani2011PRL,Catelani2011PRB,Glazman2021}. 
	Common QP-induced error mechanisms include frequency shifts and relaxation~\cite{Martinis2009QPEnergyDecay,Catelani2011PRL,Catelani2011PRB}, parity switching and non-Poissonian jumps~\cite{Catelani2014Parity,Riste2013Parity,Serniak2018HotQPs,Serniak2019,Diamond2022ParityMechanisms}, loss and dephasing in protected modes~\cite{Pop2014Suppression,Vool2014FluxoniumQP,Spilla2015FluxoniumDephasing}, and correlated errors triggered by ionizing radiation and pair-breaking photons~\cite{Wilen2021Correlated,Liu2024PairBreakingPhotons,VDKurilovich2025, mcewen_resolving_2022}.
	Several complementary strategies have been developed to mitigate quasiparticle poisoning in superconducting circuits. Superconducting-gap engineering can suppress the tunneling of pre-existing quasiparticles and protect qubit arrays against burst events~\cite{Sun2012GapEngineering,McEwen2024GapEngineering}, normal-metal traps reduce the mobile quasiparticle population~\cite{Riwar2016Traps}, and phonon-downconversion structures limit secondary pair breaking by energetic phonons~\cite{Iaia2022PhononDownconversion}. Periodic driving, however, can partially circumvent the protection provided by static gap engineering because absorption of drive photons opens inelastic tunneling channels for existing quasiparticles and can also generate new quasiparticle pairs~\cite{Kishmar2025DrivenQP,Chowdhury2025DrivenQPTheory}. These sideband-assisted processes, whose transition amplitudes additionally acquire Floquet micromotion, therefore require a treatment tailored to the driven circuit~\cite{Kishmar2025DrivenQP,Chowdhury2025DrivenQPTheory}.
	
	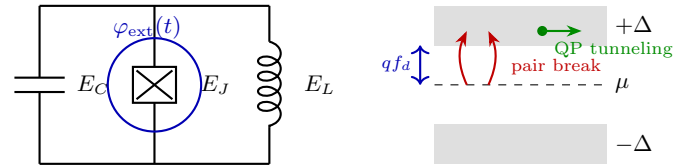
\begin{figure}[htb!]
		\centering
		\begin{tikzpicture}[
			scale=0.95, transform shape,
			every node/.style={font=\small},
			wire/.style={thick},
			ind/.style={thick,decorate,decoration={coil,aspect=0.7,
					segment length=2mm,amplitude=1.5mm}},
			jj/.style={thick}
			]
			\draw[wire]    (0,0) -- (0,1.0);
			\draw[wire]    (-0.35,1.0) -- (0.35,1.0);
			\draw[wire]    (-0.35,1.2) -- (0.35,1.2);
			\draw[wire]    (0,1.2) -- (0,2.2);
			\node at (0.75,1.1) {$\Ec$};
			\draw[jj]     (1.6,0) -- (1.6,0.85);
			\draw[jj]     (1.3,0.85) rectangle (1.9,1.35);
			\draw[jj]     (1.35,0.90) -- (1.85,1.30);
			\draw[jj]     (1.85,0.90) -- (1.35,1.30);
			\draw[jj]     (1.6,1.35) -- (1.6,2.2);
			\node at (2.45,1.1) {$\EJ$};
			\draw[ind]    (3.2,1.7) -- (3.2,0.5);
			\draw[thick]  (3.2,0) -- (3.2,0.5);
			\draw[thick]  (3.2,1.7) -- (3.2,2.2);
			\node at (3.9,1.1) {$\EL$};
			\draw[thick]  (0,2.2) -- (3.2,2.2);
			\draw[thick]  (0,0)   -- (3.2,0);
			\draw[-{Stealth[length=2.2mm]},thick,blue!70!black]
			(1.6,1.1) ++(0,0.0) circle (0.65);
			\node[blue!70!black] at (1.5,1.9) {$\phiext(t)$};
			\begin{scope}[xshift=5.6cm,yshift=0cm]
				\fill[gray!25] (-0.1,0.0) rectangle (2.3,0.55);
				\fill[gray!25] (-0.1,1.65) rectangle (2.3,2.20);
				\node[right] at (2.3,0.275) {$-\Delta$};
				\node[right] at (2.3,1.925) {$+\Delta$};
				\draw[dashed] (-0.1,1.10) -- (2.3,1.10);
				\node[right] at (2.3,1.10) {$\mu$};
				\draw[-{Stealth[length=2.0mm]},thick,red!75!black]
				(0.35,1.10) to[bend left=25] (0.35,1.85);
				\draw[-{Stealth[length=2.0mm]},thick,red!75!black]
				(0.65,1.10) to[bend right=25] (0.65,1.85);
				\node[red!75!black,below right=-1pt and 0pt] at (0.85,1.55)
				{\scriptsize pair break};
				\fill[green!50!black] (1.40,1.85) circle (0.07);
				\draw[-{Stealth[length=2.0mm]},thick,green!55!black]
				(1.40,1.85) -- (1.95,1.85);
				\node[green!55!black,below right=-1pt and 0pt] at (1.45,1.78)
				{\scriptsize QP tunneling};
				\draw[<->,thick,blue!70!black]
				(-0.30,1.10) -- (-0.30,1.65);
				\node[blue!70!black,left] at (-0.30,1.40)
				{\scriptsize $q f_d$};
			\end{scope}
		\end{tikzpicture}
		\caption{Schematic of a driven fluxonium circuit (left) and the two quasiparticle (QP) poisoning channels (right) considered in this work. 
			\emph{Left:} Circuit comprising a capacitor with charging energy \(E_C\), a small junction with Josephson energy \(E_J\), and a superinductance with inductive energy \(E_L\), threaded by a time-dependent external flux $\phiext(t)$. 
			\emph{Right:} Density of states with gap $\Delta$ in the superconducting junction with a chemical potential $\mu$, showing drive-assisted Cooper-pair breaking (red) and tunneling of a pre-existing nonequilibrium QP (green).}
		\label{fig:schematic}
	\end{figure}
	
	The goal of the paper is to formulate QP poisoning in driven fluxonium in the same irrotational gauge used to derive frozonium~\cite{You2019,Lewellen2025Frozonium}, so that the tunneling operator and the driven circuit Hamiltonian are referenced to the same junction phase. 
	We separate drive-assisted Cooper-pair breaking from the tunneling of pre-existing nonequilibrium QPs and track both BCS coherence-factor channels associated with
	$\cos(\phiJ/2)$ and $\sin(\phiJ/2)$ (see below for the definition of the time-dependent junction phase, $\phiJ$).
	We find that the pair-breaking transition rate is dominated by two effects: the $2\Delta /n$ gap-breaking threshold ($\Delta$ is the pairing gap, and $n=1,2,\dots$), which dominates at high frequencies, and the drive-induced multiphoton resonance between the fluxonium states, which dominates at low frequency.
	At the same time, the tunneling transition rate contains features {organized} by the latter, but is insensitive to the former.
	However, as we will demonstrate below, the location of these multiphoton resonances can be predicted accurately by extending the Floquet-Magnus expansion.

	\textit{Driven fluxonium.---} We consider a single-junction fluxonium loop with charging energy $\Ec$, Josephson energy $\EJ$, inductive energy $\EL$, and dynamical phase $\hat{\varphi}$ and number $\hat{n}$ obeying $[\hat{\varphi},\hat{n}]=i$. Following Ref.~\cite{Lewellen2025Frozonium}, we work in the irrotational gauge of Ref.~\cite{You2019}, in which the time-dependent flux drive is implemented as a coordinated pair of charge- and flux-like drives. After going into the co-moving frame with a suitable gauge transformation [see the End Matter (EM) for details], we obtain the following fluxonium Hamiltonian 
	\begin{equation}
		\begin{aligned}
			\Hmov(t)&=4\Ec\,\hat{n}^{2}+\frac{\EL}{2}\hat{\varphi}^{2}
			-\EJ\cos\phiJ(t),\\
			\phiJ(t)&\equiv\hat{\varphi}-\phiext-\Theta(t).
		\end{aligned}
		\label{eq:Hmov}
	\end{equation}
	Throughout we set $\phiext=\pi$ (half-flux point) for first-order flux-noise insensitivity~\cite{Manucharyan2009Fluxonium,Nguyen2019HighCoherence}. The phase drive is $\Theta(t)=\phi_{\rm ac}\sin(\omega t)$, with frequency $f_d=\omega/(2\pi)$ and amplitude $\phi_{\rm ac}$.
	
	In the high-frequency limit, the dynamics of the system can be approximated by the Floquet-Magnus expansion, whose leading-order effective Hamiltonian is 
	\begin{equation}
		\begin{aligned}
			&\calH_{\rm M}^{(0)}(\phi_{\rm ac})
			=\frac{1}{T}\int_0^T dt\,\Hmov(t)\\
			&=4\Ec\hat n^2+\frac{\EL}{2}\hat\varphi^2
			-\EJ J_0(\phi_{\rm ac})\cos(\hat\varphi-\phiext)\,,
		\end{aligned}
		\label{eq:HMagnus}
	\end{equation} where $J_0(...)$ denotes the Bessel function. 
	At the zeros of $J_{0}(...)$, Eq.~\eqref{eq:HMagnus} becomes a quadratic effective Hamiltonian
	$\Hquad=4\Ec\hat{n}^{2}+(\EL/2)\hat{\varphi}^{2}$ with fundamental frequency
	\begin{equation}
		\nu = \sqrt{8\Ec\EL}/h,
		\label{eq:nu}
	\end{equation}
	and residual nonlinearity suppressed in powers of $1/\omega$~\cite{Lewellen2025Frozonium}. 
	The first freezing amplitude for a monochromatic sine drive is $\phi^{\star}_{\rm ac}\approx2.4048$. For the rest of the paper, the fluxonium parameters are $\EJ/h=4.88\,\mathrm{GHz}$, $\Ec/h=1.09\,\mathrm{GHz}$, and $\EL/h=0.56\,\mathrm{GHz}$, giving $\nu\simeq2.21\,\mathrm{GHz}$.  The superconducting gap is $f_\Delta=\Delta/h=45\,\mathrm{GHz}$, appropriate for aluminum~\cite{McEwen2024GapEngineering}.

	\textit{Floquet quasiparticle transition rates.---} The Josephson coupling in the fluxonium Hamiltonian introduced above arises microscopically from the tunneling term of electrons, given by 
	\begin{equation}
		\hat{H}_{T}(t)=
		\sum_{kk'\sigma}\tau_{kk'}\,
		e^{i\phiJ(t)/2}c^{\dagger}_{Rk\sigma}c^{\phantom\dagger}_{Lk'\sigma}
		+\mathrm{h.c.}\,, \label{eq:HT}
	\end{equation} which is written in the same gauge as Eq.~\eqref{eq:Hmov}. 
	Here $\tau_{kk'}$ is the single-electron tunneling amplitude. 
	In calculations below, we use a momentum-independent form $\tau_{kk'}\to\tau$, which is related to $E_J$ by $E_J=\pi^2\nu_0^2\Delta|\tau|^2$~\cite{Catelani2011PRB,Glazman2021}. Here, $\nu_0$ is the density of states at the Fermi level per spin.
	The nondissipative virtual effects of the tunneling term above lead to the Josephson coupling term $E_J$ in Eq.~\eqref{eq:Hmov} via standard second-order perturbation theory.

	We calculate the pair-generation rate and pre-existing-quasiparticle tunneling rate due to Eq.~\eqref{eq:HT} using Floquet Fermi's Golden Rule (FFGR)~\cite{Kishmar2025DrivenQP,Chowdhury2025DrivenQPTheory}, with the derivations deferred to EM. The matrix elements used in FFGR are computed by exact diagonalization (ED) implemented using QuSpin~\cite{PWeinberg2017,PWeinberg2019}. 
	We denote the total state-changing rate out of Floquet branch $\alpha$ by $\Gamma_\alpha^X=\sum_{\beta\neq\alpha}\Gamma_{\alpha\to\beta}^X$, where $X\in\{\mathrm{pair},\mathrm{tunnel}\}$.
	The Floquet branch labels
	are defined relative to an amplitude-continuous eigenbasis of the Magnus Hamiltonian in
	Eq.~\eqref{eq:HMagnus}, which is adiabatically connected to the undriven fluxonium eigenbasis~\cite{supp}.
	At each $(f_d,\phi_{\rm ac})$, the Floquet states computed from ED are assigned to this reference by the one-to-one permutation that maximizes their total overlap~\cite{supp}.
	
	\begin{figure*}[htb!]
		\centering
			\includegraphics[width=\textwidth]{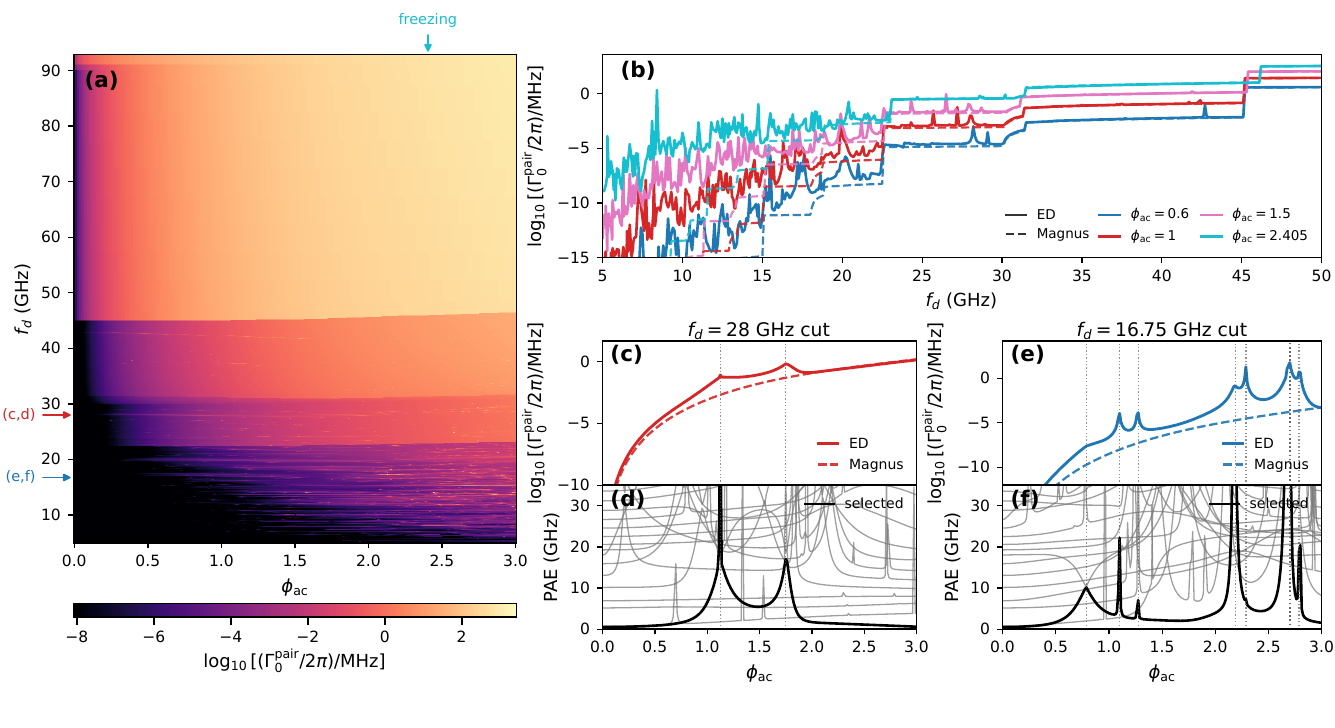}
		\caption{Pair-generation rate landscape for fluxonium branch $0$, which denotes the Floquet eigenstate assigned to the Magnus ground state by the one-to-one overlap procedure. 
			(a) Quasiparticle generation rate $\Gamma^\text{pair}_0$ as a function of phase-drive amplitude, $\phi_{\rm ac}$, and drive frequency, $f_d$, plotted on a log scale. 
			The top-right arrow marks the location of the freezing point $\phi_{\rm ac}\simeq 2.405$. 
			(b) Fixed-$\phi_{\rm ac}$ cuts compare $\Gamma^\text{pair}_0$ computed from ED (solid lines) with the Magnus reference (dashed lines); see \cite{supp} for more details.
			The reference captures the high-frequency magnitude and the discontinuous openings at the $2\Delta/n$
			thresholds, while visible deviations appear below about $20~{\rm GHz}$.
			At fixed $f_d=28~{\rm GHz}$, a rate spike (panel c) absent from the static-Magnus (dashed line)
			background is correlated with a hybridization feature in the period-averaged-energy (PAE) spectrum; see panel (d).
			Panels (e) and (f) show the correspondence between a rate spike and a resonance at the lower frequency $f_d=16.75~{\rm GHz}$. 
			As the frequency is lowered, the density of resonances increases. In (d) and (f), the PAE of branch $0$ is highlighted in black, and the PAEs of other levels are plotted in gray. 
		}
		\label{fig:pair-rate-landscape}
	\end{figure*}
	
	\begin{figure*}[htb!]
		\centering 
		\includegraphics[width=\textwidth]{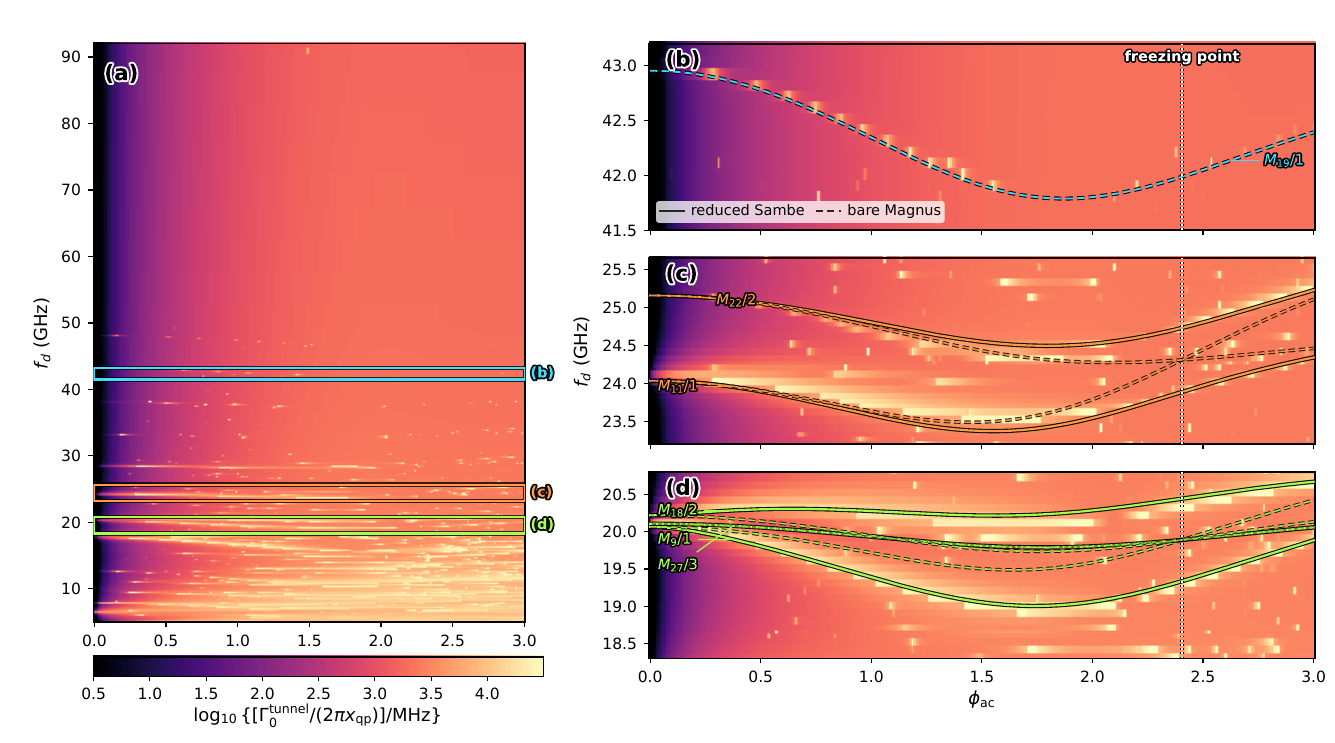}
		\caption{Quasiparticle tunneling rate landscape for fluxonium branch $0$, tracked using the same numerical procedure as in Fig.~\ref{fig:pair-rate-landscape}. 
			(a) $\Gamma^\text{tunnel}_0/x_\text{qp}$ as a function of phase-drive amplitude, $\phi_{\rm ac}$, and drive frequency, $f_d$, plotted on a log scale.  
			In addition to the isolated resonances studied in Fig.~\ref{fig:pair-rate-landscape}, $\Gamma_0^\text{tunnel}$ contains connected resonance curves whose structure is organized by the Magnus spectrum and hybridization among resonant channels. 
			(b) A one-photon resonance whose location is accurately predicted by the bare Magnus condition $E^M_{19}-E_0^M=hf_d$. The dashed $M_{19}/1$ locus follows the bright rate ridge. 
			(c) The bare $M_{11}/1$ and $M_{22}/2$ loci meet at the freezing point as a consequence of frozonium harmonicity. 
			Hybridization between these two resonant channels produces the solid avoided-crossing loci with $M_0$. 
			(d) The corresponding three-channel hybridization involving $M_9/1$, $M_{18}/2$, and $M_{27}/3$, which become degenerate at the freezing point. }\label{fig:tunneling}
	\end{figure*}
	
	Fig.~\ref{fig:pair-rate-landscape} shows $\Gamma^\text{pair}_0$, the pair-breaking state-changing rate of branch $0$, which is assigned to the Magnus ground state by the procedure described above. Panel (a) shows $\Gamma_0^\text{pair}$ as a function of $f_d$ and $\phi_{\rm ac}$. As a function of $f_d$, $\Gamma_0^\text{pair}$ shows a series of threshold features whose frequencies lie approximately at $2f_\Delta/n$, corresponding to the nonlinear processes where the system absorbs $n$ photons and creates a pair of Bogoliubov quasiparticles. The threshold frequency disperses weakly with $\phi_{\rm ac}$, which can be attributed to the dispersion of the Floquet quasienergies as a function of $\phi_{\rm ac}$. In panel (b) we show several vertical cuts of $\Gamma_0^\text{pair}$ at different $\phi_{\rm ac}$ as solid lines, and contrast them with transition rates computed using the Magnus Hamiltonian in Eq.~\eqref{eq:HMagnus}~\cite{supp}. We see that the Magnus reference correctly captures $\Gamma_0^\text{pair}$ at high $f_d$, down to $f_d\approx {20~{\rm GHz}}$, which is well above the frozonium frequency $\nu$ as well as other circuit parameters.
	
	We also note that as far as the pair generation rate is concerned, the freezing point of fluxonium does not bring special suppression of $\Gamma_0^\text{pair}$, as labeled explicitly in panels (b)--(d).
	
	\textit{Resonances and transition-rate hotspots.---} In the low $f_d$ region of Fig.~\ref{fig:pair-rate-landscape}, we see a number of spike-like features that also show clear deviations from the Magnus prediction for $\Gamma^\text{pair}_0$ [see panels (c) and (e) of Fig.~\ref{fig:pair-rate-landscape}]. 
	We now demonstrate that these features are associated with resonances, namely, strong hybridization of Floquet eigenstates when the corresponding Magnus eigenstates are connected by a multiphoton drive term~\cite{Eckardt2015,Bukov2015}. 
	To identify resonances independently of the rate features, we use the period-averaged energy (PAE) defined as follows~\cite{Ketzmerick2010FloquetStatistics,PMSchindler2024,Lewellen2025Frozonium}: 
	\begin{equation}
		E_{\alpha}^{\rm PAE}
		=\frac{1}{T}\int_0^T dt\,
		\langle u_{\alpha}(t)|\Hmov(t)|u_{\alpha}(t)\rangle \,.
		\label{eq:PAE}
	\end{equation}  Here $\ket{u_\alpha(t)}$ is the Floquet eigenvector of Eq.~\eqref{eq:Hmov}. 
	Unlike Floquet quasienergy, PAE is gauge invariant under Floquet gauge transformations, and therefore serves as an absolute sorting device of the Floquet eigenstates \cite{PMSchindler2024}. 
	In the high-frequency regime, PAE approaches the Magnus eigenvalues of Eq.~\eqref{eq:HMagnus} up to $1/\omega$ corrections, and resonances appear as crossing-like features in PAE \cite{PMSchindler2024,Lewellen2025Frozonium}. 
	Examples of resonant features in PAE are shown in panels (d) and (f) of Fig.~\ref{fig:pair-rate-landscape}, where we have plotted the dispersion of PAEs with $\phi_{\rm ac}$ at fixed $f_d$. 
	The location of the resonances can be predicted within the Magnus expansion, i.e., they occur when the Magnus energies are connected by a multiphoton resonance $E_\alpha^M-E_\beta^M=nhf_d$ \cite{Lewellen2025Frozonium}. 
	As we illustrate in panels (c) and (e), the local spikes in $\Gamma^\text{pair}_0$ can be directly matched with the resonance features of the PAE of branch $0$ in panels (d) and (f), respectively. Furthermore, the density of these isolated resonances increases with decreasing $f_d$. 
	Thus, lowering \(f_d\) has two competing effects on the quasiparticle-generation rate: it suppresses direct multiphoton pair breaking but increases the density of Floquet resonances. 
	The observation of these isolated spikes establishes a resonance-assisted pair-breaking mechanism: multiphoton hybridization gives branch $0$ access to high-energy circuit states whose energy exceeds the pair-breaking threshold. 
	As these resonances proliferate at lower \(f_d\), their associated enhancements form an increasingly dense rate pattern.

	\textit{Photon-assisted quasiparticle tunneling.---} We now turn to the quasiparticle-tunneling rate.
	We computed the state-changing transition rate $\Gamma_\alpha^\text{tunnel}=\sum_{\beta\neq \alpha} \Gamma_{\alpha\to\beta}^\text{tunnel}$ and the result is shown in Fig.~\ref{fig:tunneling}. The tunneling rates are reported per unit dimensionless quasiparticle density $\xqp=n_{\rm qp}/(2\nu_0\Delta)$, where $n_{\rm qp}$ is the density of stray quasiparticles~\cite{Catelani2011PRB}. 
	We focus again on branch $0$, whose tunneling rate landscape is shown in panel (a). 
	Unlike $\Gamma_0^\text{pair}$, $\Gamma_0^\text{tunnel}$ does not contain threshold behavior at $2f_\Delta/n$, but instead contains a number of resonant spikes, including both isolated resonances, discussed above, and continuous resonances. 
	As we explain below, the continuous features in $\Gamma^\text{tunnel}_0$ are organized by multiphoton-resonance crossings inherited from the harmonic Magnus spectrum near the freezing point, highlighting the broader relevance of frozonium phenomenology to quasiparticle-based phenomena.
	
	\textit{Characterizing continuous resonances.---} Fig.~\ref{fig:tunneling}(b) shows a simple resonance that is predicted accurately by the leading-order Magnus spectrum. 
	We use $M_n/p$ to denote a $p$-photon resonance between the Magnus levels $M_n$ and $M_0$, defined by
	\begin{equation}\label{eq:22}
		E^M_n(\phi_{\rm ac})-E_0^M(\phi_{\rm ac})=p h f_d\,.
	\end{equation}
	The $M_{19}/1$ locus closely follows the bright ridge; its apparent discontinuity results from the finite $0.1\,{\rm GHz}$ frequency resolution.
	More complex structures arise when several bare resonance loci meet. 
	As noted previously, at the frozonium freezing point, $J_0(\phi_{\rm ac}^\star)=0$, and the leading-order Magnus Hamiltonian Eq.~\eqref{eq:HMagnus} becomes harmonic \cite{Lewellen2025Frozonium}. 
	Consequently, the family $M_n/1,M_{2n}/2,M_{3n}/3,\ldots$ becomes degenerate; coupling among these resonant channels converts the bare crossings into avoided crossings, producing the two dressed branches in Fig.~\ref{fig:tunneling}(c), and the three dressed branches in Fig.~\ref{fig:tunneling}(d), respectively. 
	A detailed formulation of the avoided crossing is presented in the End Matter using the Sambe formalism.
	
	\textit{Implications for experiments.---} We finally discuss some consequences of the above results for an aluminum-based device with the circuit parameters introduced above. 
	We choose a representative value of the quasiparticle concentration $\xqp=10^{-6}$ to model a device away from ionizing radiation, and fix the drive
	amplitude at the freezing point $\phi_{\rm ac}=\phi^\star_{\rm ac}$.  
	For the small junction of aluminum fluxonium devices, reported or inferred bounds range from $\xqp\lesssim5\times10^{-9}$ to $\xqp\sim1.2\times10^{-5}$~\cite{Pop2014Suppression,Grunhaupt2019GranularAl,Somoroff2023Millisecond,Atanasova2025PostCavity,Watanabe2025Fluorescence,Ateshian2025Relaxation,Azar2026Relaxation}.  
	During radiation-induced bursts the instantaneous density can be much larger: granular-aluminum resonators used $\delta\xqp=5\times10^{-5}$ as the burst threshold and showed individual excursions reaching several times $10^{-4}$~\cite{Cardani2021}.
	
	\begin{figure}[htb!]
		\centering
		\includegraphics[width=\columnwidth]{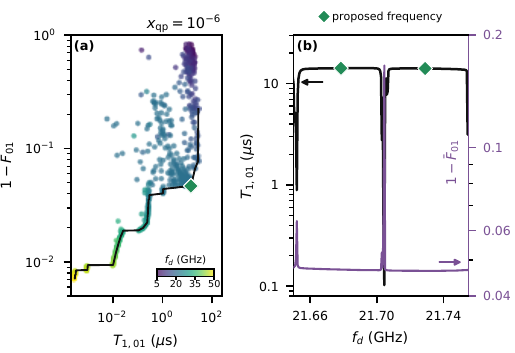}
		\caption{Freezing-point operating tradeoff for an aluminum-based circuit with $\xqp=10^{-6}$. 
			(a) Pointwise drive-frequency scan in the $(T_{1,01},1-\bar F_{01})$ plane. {Two proposed drive frequencies are highlighted.} 
			(b) Fine scan around $21.7\,\mathrm{GHz}$ reveals two operating frequency intervals separated by a resonance.  
			The black and purple curves show $T_{1,01}$ and $1-\bar F_{01}$ on the left and right axes, respectively.  
			The narrow feature near $21.704\,\mathrm{GHz}$ is the nearest resolved resonance.  
			The left and right highlighted frequencies are $f_d=21.6785\,\mathrm{GHz}$ and $21.7290\,\mathrm{GHz}$, respectively, placing them $25.5\,\mathrm{MHz}$ and $24.5\,\mathrm{MHz}$ from the nearest resonance.}
		\label{fig:experiment-pareto}
	\end{figure}
	
	To quantify the approach to the harmonic limit in the manifold of the lowest two states, we define $\bar F_{01}=(F_0+F_1)/2$, where $F_n=|\langle u_n|M_n\rangle|^2$ is the fidelity of the Floquet eigenstate with the corresponding harmonic Magnus state. 
	We combine the state-resolved population-loss rates through $T_{1,01}^{-1}=(\Gamma_0+\Gamma_1)/2$, with $\Gamma_n=\Gamma_n^{\rm pair}+\Gamma_n^{\rm tunnel}$.
	Fig.~\ref{fig:experiment-pareto} displays the resulting lifetime-fidelity tradeoff as the drive frequency is varied.  
	The solid line in Fig.~\ref{fig:experiment-pareto}(a) identifies frequencies for which neither $T_{1,01}$ nor $\bar F_{01}$ can be improved without degrading the other. 
	To identify frequencies that balance the near-harmonicity produced by dynamical freezing against quasiparticle-poisoning rates, we performed a detailed scan in panel (b) near $21.7\,\mathrm{GHz}$; a narrow resonance around $21.704\,\mathrm{GHz}$ separates two broad operating intervals. 
	Choosing either proposed frequency $f_d=21.68\,\mathrm{GHz}$ or $21.73\,\mathrm{GHz}$ yields $T_{1,01}\simeq{14}\,\mu\mathrm{s}$ and $\bar F_{01}\simeq0.953$, while placing the drive approximately $25\,\mathrm{MHz}$ away from the nearest resolved resonance in either case. 
	
	\textit{Discussion and Outlook.-}
	The lifetime and fidelity estimates above reveal a practical operating-window tradeoff for frozonium: increasing the drive frequency improves convergence toward the harmonic-oscillator limit but enhances drive-induced pair generation, whereas lowering the frequency leads to a growing density of Floquet resonances. Higher-gap junction platforms can retain conventional microwave circuit parameters while shifting the pair-breaking thresholds upward~\cite{Anferov2024,Wang2026AllNitride}, potentially widening the frequency range over which high oscillator fidelity can be achieved without prohibitive quasiparticle loss. Our results therefore identify the superconducting gap and resonance-aware drive selection as coupled design parameters for realizing frozonium experimentally.

    A related question is how the circuit parameters \(E_L\) and \(E_J\) affect the above picture. Decreasing \(E_L\) lowers the frozonium frequency \(\nu\), thereby shifting the resonance structure to lower frequencies. Near the freezing point, varying \(E_J\) does not affect the frozonium states at leading order in the Magnus expansion, although the quasiparticle-poisoning rates studied here scale linearly with \(E_J\). Another important direction is therefore to determine how far \(E_L\) and \(E_J\) can be reduced before other loss mechanisms become dominant.

	An important future direction is to determine how drive-generated quasiparticles affect $T_2$, beyond their contribution to $T_1$. Since the junction parameters are sensitive to the quasiparticle population near the junction, it is important to model the transport of the generated quasiparticles around the junction. In the current work, we have assumed that generated quasiparticles are promptly removed from the junction. Modeling this transport requires a kinetic-equation formalism that incorporates quasiparticle diffusion, relaxation, and trapping, and connects the resulting local quasiparticle distribution to dephasing~\cite{Wang2014QPDynamics,Riwar2016Traps,Spilla2015FluxoniumDephasing}.

	We also briefly discuss how our results differ from previous studies of transmons~\cite{Houzet2019PhotonAssisted,Chowdhury2025DrivenQPTheory}. One notable difference is that, in fluxonium, resonances play a dominant role in determining the quasiparticle-poisoning rates at low drive frequencies, when $f_d$ becomes comparable to the frozonium frequency $\nu$. This feature is not prominent in Refs.~\cite{Houzet2019PhotonAssisted,Chowdhury2025DrivenQPTheory}. The difference may be attributed to the distinct role of the offset charge $n_g$. Fluxonium is insensitive to $n_g$~\cite{Manucharyan2009Fluxonium,Koch2009}, enabling a self-contained treatment of its driven Hilbert space. In transmons, however, both quasiparticle generation and tunneling change charge parity, thereby connecting the two charge sectors of the circuit Hilbert space. A Floquet treatment at fixed $n_g$ does not capture possible resonances between these sectors as $n_g$ fluctuates. Given the recently observed sensitivity of multiphoton resonances in transmons to $n_g$~\cite{Fechant2025OffsetCharge}, it would be interesting to investigate whether analogous resonance-enhanced quasiparticle poisoning occurs in transmons.

	\begin{acknowledgments}
		We thank K. J. Lewellen, R. Mukherjee, S. Roy and V. Fatemi for useful discussions and collaboration on related work. This material is based on research sponsored by  the Air Force Office of Scientific Research under agreement number FA9550-26-1-B088. The U.S. Government is authorized to reproduce and distribute reprints for Governmental purposes notwithstanding any copyright notation thereon. We disclose the use of ChatGPT Codex (GPT 5.5 and 5.6) for coding, plotting, and preparation of the manuscript, and the results are verified by the authors.
	\end{acknowledgments}

	\bibliography{references}

	\clearpage
	
	\begin{center}
		{\large\bfseries End Matter}
	\end{center}

	\textit{Realizing phase-driven fluxonium from coordinated charge and flux drives.---} In this section, we review how to generate the phase drive Eq.~\eqref{eq:Hmov} from a coordinated charge + flux drive \cite{Lewellen2025Frozonium, You2019}.  We start from the Hamiltonian
	\begin{equation}
		\mathcal{H}(t) = \Hflux - f(t)\,\hat{n} + g(t)\,\hat{\varphi},
		\qquad g(t) = \EL\,\Theta(t),
		\label{eq:Hlab}
	\end{equation}
	with $\Theta(t)\equiv\int_{0}^{t}\!dt'\,f(t')$. 
	Here, 
	\begin{equation}
		\Hflux = 4\Ec\,\hat{n}^{2} + \frac{\EL}{2}\hat{\varphi}^{2}
		- \EJ\cos\big(\hat{\varphi}-\phiext\big),
		\label{eq:Hflux}
	\end{equation}
	with $\phiext$ the static external flux through the loop. 
	For concreteness we focus on a single-tone sinusoidal drive, $f(t)=A\cos(\omega t)$, so that $\Theta(t)=\phi_{\rm ac}\sin(\omega t)$ with $\phi_{\rm ac}\equiv A/\omega$ the drive phase amplitude and $f_{d}\equiv\omega/(2\pi)$. 
	Analogous results for the triangular drive of Ref.~\cite{Lewellen2025Frozonium} can be obtained by replacing the single Bessel sideband structure with the corresponding Fourier decomposition.
	
	{Defining the co-moving state as $|\psi_{\rm mov}(t)\rangle=W(t)|\psi_{\rm lab}(t)\rangle$, with $W(t)=\exp[-i\Theta(t)\,\hat{n}]$, gives the gauge transformation $\Hmov(t)=W(t)\calH(t)W^\dagger(t)+i\dot{W}(t)W^\dagger(t)$, up to the purely time-dependent scalar specified below, and trades the two lab-frame drives for a single time-dependent shift of the junction phase~\cite{Lewellen2025Frozonium}.} 
	{After dropping the purely time-dependent scalar $-\EL\Theta^2(t)/2$, the Hamiltonian reduces to Eq.~\eqref{eq:Hmov} in the main text.}
	
	\textit{Floquet Fermi's Golden Rule.---}
	The dissipative effects of Eq.~\eqref{eq:HT} are quasiparticle generation and photon-assisted quasiparticle tunneling~\cite{Catelani2011PRB,Houzet2019PhotonAssisted,Kishmar2025DrivenQP,Chowdhury2025DrivenQPTheory}, which can be described by the two perturbations
	$\hat{H}_T\rightarrow \hat{V}_{\mathrm{gen}}+\hat{V}_{\mathrm{tun}}$. The explicit forms below follow from substituting the Bogoliubov quasiparticle representation of the electron operators into Eq.~\eqref{eq:HT}: 
	\begin{align}
		\hat V_{\mathrm{gen}}(t)
		={}&{\sum_{\sigma}\int_{\bm{k}}\int_{\bm{k}'}}\tau_{kk'}\big[A^{+}_{kk'}\OC(t)+iA^{-}_{kk'}\OS(t)\big]\notag\\[-0.3em]
		&\quad{}\times\gamma^{\dagger}_{Rk\sigma}\gamma^{\dagger}_{Lk'\bar\sigma}+\mathrm{h.c.},\label{eq:Vgen} \\
		\hat V_{\mathrm{tun}}(t)
		={}&{\sum_{\sigma}\int_{\bm{k}}\int_{\bm{k}'}}\tau_{kk'}\big[B^{-}_{kk'}\OC(t)+iB^{+}_{kk'}\OS(t)\big]\notag\\[-0.3em]
		&\quad{}\times\gamma^{\dagger}_{Rk\sigma}\gamma^{\phantom\dagger}_{Lk'\sigma}+\mathrm{h.c.}\,, \label{eq:Vtun}
	\end{align}
	where \({\int_{\bm{k}}\equiv\int \mathrm{d}^{3}k/(2\pi)^{3}}\).
	Here $\gamma$ denotes a Bogoliubov quasiparticle, and the amplitudes
	\begin{subequations}
		\begin{align}
			A^{\pm}_{kk'}&=u_{Rk}v_{Lk'}\pm v_{Rk}u_{Lk'},\\
			B^{\pm}_{kk'}&=u_{Rk}u_{Lk'}\pm v_{Rk}v_{Lk'} \,,
		\end{align}
		\label{eq:coherence}
	\end{subequations}
	are expressed in terms of $u,v$, the usual BCS coherence factors, with $R,L$ representing the two sides of the Josephson junction and $k$ the quasiparticle momentum. 
	These amplitudes couple to the  circuit operators, 
	\begin{equation}
		\OC(t)=\cos\!\left[\frac{\phiJ(t)}{2}\right],\qquad
		\OS(t)=\sin\!\left[\frac{\phiJ(t)}{2}\right].
		\label{eq:Ocs}
	\end{equation}
	Here the factor of $1/2$ arises since $\hat{\varphi}$ encodes the phase of the condensate, and $\hat{\varphi}_J$ is defined in Eq.~\eqref{eq:Hmov}.
	Eqs.~\eqref{eq:Vgen} and \eqref{eq:Vtun} are the starting point for calculating transition rates using Floquet Fermi's Golden Rule.  
	
	\textit{Pair-breaking transitions.---} For Cooper-pair-breaking transitions, we obtain 
	\begin{align}
		\Gamma^{\mathrm{pair}}_{\alpha\to\beta}
		=\frac{16\EJ}{h}\sum_{q\in \mathbb{Z}}
		\bigg[
		&\big|M^{(q)}_{c,\beta\alpha}\big|^{2}
		{S_{\mathrm{gen}}^{c}}\!\left(\frac{f_{\alpha\beta}^{(q)}}{f_{\Delta}}\right)
		\notag\\
		&+
		\big|M^{(q)}_{s,\beta\alpha}\big|^{2}
		{S_{\mathrm{gen}}^{s}}\!\left(\frac{f_{\alpha\beta}^{(q)}}{f_{\Delta}}\right)
		\bigg],
		\label{eq:Gpair}
	\end{align}
	where $f_{\Delta}=\Delta/h$~\cite{Catelani2011PRL,Catelani2011PRB,Glazman2021}, and the  matrix element is
	\begin{equation}
		M^{(q)}_{p,\beta\alpha}=
		\frac{1}{T}\int_{0}^{T}\!dt\,e^{+iq\omega t}
		\langle u_{\beta}(t)|\hat{O}_{p}(t)|u_{\alpha}(t)\rangle,~~p=c,s.
		\label{eq:Mq}
	\end{equation}  
	Here $\alpha,\beta$ denote Floquet eigenstates of Eq.~\eqref{eq:Hmov}, where the Floquet eigenvector is $|\psi_{\alpha}(t)\rangle=e^{-i\epsilon_{\alpha}t/\hbar}|u_{\alpha}(t)\rangle$, with
	$|u_{\alpha}(t+T)\rangle=|u_{\alpha}(t)\rangle$ and $T=1/f_{d}$, and the quasienergy is $\epsilon_\alpha$.
	The sum over $q$ runs over the Floquet sidebands, which can be understood as the number of photons absorbed (emitted) during the transition. 
	The corresponding frequency delivered to the quasiparticle sector is
	\begin{equation}
		f_{\alpha\beta}^{(q)}
		=\frac{\epsilon_{\alpha}-\epsilon_{\beta}}{h}+qf_{d}.
		\label{eq:fq}
	\end{equation}
	Positive $f_{\alpha\beta}^{(q)}$ means that the quasiparticle sector absorbs energy. 
	We note that only the sum over $q$ in Eq.~\eqref{eq:Gpair} is gauge invariant, whereas individual terms can be shifted under Floquet gauge transformations, $\epsilon_\alpha\to\epsilon_\alpha+h f_d$, $\ket{u_\alpha(t)}\to e^{i\omega_d t}\ket{u_\alpha(t)}$. 
	In this work, we computed the Floquet eigenvectors $\ket{u_\alpha(t)}$, quasienergies $\epsilon_\alpha$, and transition amplitudes $M^{(q)}_{p,\beta\alpha}$ using ED implemented using QuSpin package~\cite{PWeinberg2017,PWeinberg2019}. {Further numerical details are given in the Supplemental Material~\cite{supp}.} Some coding tasks were performed using Codex and benchmarked against previous results \cite{Lewellen2025Frozonium}. 
	
	The form factors ${S_{\mathrm{gen}}^{c,s}}(z)$ in Eq.~\eqref{eq:Gpair} are the sums over two
	real-quasiparticle final states weighted by the coherence factors in
	Eq.~\eqref{eq:coherence},
	\begin{align}
		{S_{\mathrm{gen}}^{c,s}}(z)&=
		{\frac{1}{2(\nu_{0}\Delta)^{2}}\int_{\bm{k}}\int_{\bm{k}'}}\notag\\[-0.25ex]
		&\quad\times
		\left|u_{Rk}v_{Lk'}\pm v_{Rk}u_{Lk'}\right|^{2}
		\delta\!\left(z-\frac{E_{Rk}+E_{Lk'}}{\Delta}\right),
		\label{eq:Spm}
	\end{align}
	where $E_{\ell k}=(\xi_{\ell k}^{2}+\Delta^{2})^{1/2}$.  
	After assuming the density of states can be approximated by its value at the Fermi level, the above integrals can be explicitly evaluated for $z\ge 2$ as
	\begin{align}
		{S_{\mathrm{gen}}^{c}}(z)&=(z+2)E(m)-\frac{4z}{z+2}K(m),\\
		{S_{\mathrm{gen}}^{s}}(z)&=(z+2)E(m)-4K(m),
		\label{eq:SpmElliptic}
	\end{align}
	where $m=[(z-2)/(z+2)]^2$, and $K(m)$ and $E(m)$ are complete elliptic integrals of the first and second kind, respectively~\cite{Catelani2011PRL,Catelani2011PRB,Glazman2021}.
	Qualitatively, the delta function enforces energy conservation, so ${S_{\mathrm{gen}}^{c,s}}(z<2)=0$ and the \(q\)th sideband contributes to Eq.~\eqref{eq:Gpair} only when $f_{\alpha\beta}^{(q)}\ge2f_{\Delta}$.  
	At threshold, the ${S_{\mathrm{gen}}^{c}}$ coherence channel turns on sharply from the BCS edge singularity, whereas ${S_{\mathrm{gen}}^{s}}$ is coherence-factor suppressed and turns on more gently; far above threshold the two form factors become comparable as the available quasiparticle phase space grows. 
	
	For the tunneling part, Floquet Fermi's Golden Rule yields
	\begin{align}
		\Gamma^{\mathrm{tunnel}}_{\alpha\to\beta}
		={}&\sum_{q\in\mathbb Z}\bigg[
		\big|M^{(q)}_{s,\beta\alpha}\big|^{2}
		{S_{\mathrm{tun}}^{s}}\!\left(f_{\alpha\beta}^{(q)}\right)
		\notag\\[-0.25ex]
		&\qquad+\big|M^{(q)}_{c,\beta\alpha}\big|^{2}
		{S_{\mathrm{tun}}^{c}}\!\left(f_{\alpha\beta}^{(q)}\right)
		\bigg]\theta_{\mathrm H}\!\left(f_{\alpha\beta}^{(q)}\right).
		\label{eq:GqpRevised}
	\end{align}
	The matrix elements and frequency exchange are defined in Eqs.~\eqref{eq:Mq} and \eqref{eq:fq}, respectively.  
	The step function restricts the total energy transferred to the quasiparticle sector to $hf_{\alpha\beta}^{(q)}>0$.
	Particle-hole symmetry removes the interference between the sine and cosine amplitudes after carrying out the momentum sum. 
	Following the standard tunneling-Hamiltonian derivation~\cite{Catelani2011PRB,Glazman2021}, we take the same stationary occupation $f_E(E)$ in both electrodes.
	The corresponding tunneling form factors are
	\begin{align}
		&{S_{\mathrm{tun}}^{s,c}}(f)
		={}\frac{2\pi}{\hbar}{\sum_{\sigma}\int_{\bm{k}}\int_{\bm{k}'}}|\tau_{kk'}|^2 f_E(E_{Lk'})[1-f_E(E_{Rk})]
		\notag\\[-0.25ex]
		&\times\left|u_{Rk}u_{Lk'}\mathbin{\pm}v_{Rk}v_{Lk'}\right|^2 \delta(E_{Rk}-E_{Lk'}-hf)+(L\leftrightarrow R),
		\label{eq:StunMicroscopic}
	\end{align}
	where the upper sign denotes the sine channel and the lower sign denotes the
	cosine channel.
	If the occupied quasiparticles lie within a characteristic energy $\delta E_{\mathrm{qp}}$ of the gap edge and $\delta E_{\mathrm{qp}}\ll hf$ (cold-edge approximation)~\cite{Catelani2011PRB,Kishmar2025DrivenQP}, the two form factors can be approximated by
	\begin{subequations}
		\begin{align}
			{S_{\mathrm{tun}}^{s}}(f)
			&=\xqp\,\frac{8\EJ}{\pi{\hbar}}\sqrt{\frac{2f_\Delta+f}{f}},
			\label{eq:SqpSColdEdge}\\
			{S_{\mathrm{tun}}^{c}}(f)
			&=\xqp\,\frac{8\EJ}{\pi{\hbar}}\sqrt{\frac{f}{2f_\Delta+f}}.
			\label{eq:SqpCColdEdge}
		\end{align}
		\label{eq:SqpColdEdge}
	\end{subequations}
	Here the dimensionless quasiparticle density is defined by
	\begin{equation}\label{eq:xqp}
		\xqp\equiv\frac{n_{\rm qp}}{2\nu_0 \Delta}=\frac{2}{\Delta}
		\int_\Delta^\infty\!dE\,
		\frac{E}{\sqrt{E^2-\Delta^2}}f_E(E).
	\end{equation} 
	Effects of finite $\delta E_{\rm qp}$ are discussed in the Supplemental Material~\cite{supp}. Retaining a finite $\delta E_{\rm qp}$ lowers the tunneling rate; therefore, the approximate calculation used here gives a conservative lifetime estimate.

	\textit{Magnus Resonance Theory.---} 
	We now develop the reduced-Sambe description that predicts the bare and
	hybridized resonance loci in Fig.~\ref{fig:tunneling}.  We begin by expanding the co-moving Hamiltonian as
	$\Hmov(t)=\sum_{\ell}H_\ell e^{i\ell\omega t}$, where
	$H_\ell=T^{-1}\int_0^Tdt\,e^{-i\ell\omega t}\Hmov(t)$ and $H_0$ is the
	leading Magnus Hamiltonian in Eq.~\eqref{eq:HMagnus}.  Let $E_n^M$ and
	$|M_n\rangle$ denote its eigenvalues and amplitude-continuous eigenstates.
	A photon-shifted excited level becomes degenerate with $M_0$ at the bare
	$p$-photon resonance
	\begin{equation}
		E_n^M(\phi_{\rm ac})-E_0^M(\phi_{\rm ac})=phf_d.
		\label{eq:EMBareResonance}
	\end{equation}
	This condition gives the dashed $M_{19}/1$ curve in Fig.~\ref{fig:tunneling}(b).
	At the freezing amplitude, $H_0=\Hquad$ is harmonic and
	$E_n^M-E_0^M=nh\nu$.  Consequently, every member of the commensurate family
	$M_n/p,M_{2n}/(2p),\ldots$ has the same bare frequency,
	\begin{equation}
		\frac{E_{jn}^M-E_0^M}{jph}=\frac{n\nu}{p}.
		\label{eq:EMHarmonicFamily}
	\end{equation}
	The repeated intersections of bare resonance curves at freezing therefore
	follow directly from the harmonic Magnus spectrum.
	
	To include hybridization at such an intersection, we project the Floquet
	operator $\mathcal K=\Hmov(t)-i\hbar\partial_t$ into Sambe space.  For the
	family $M_n/p,M_{2n}/(2p),\ldots,M_{Ln}/(Lp)$, the reduced basis is
	$|j\rangle=|M_{jn},-jp\rangle$, where
	$|M_a,m\rangle=|M_a\rangle e^{im\omega t}$.  In frequency units, define
	\begin{align}
		C_{jk}^{(n,p)}={}&\delta_{jk}\frac{E_{jn}^M-E_0^M}{h}
		+(1-\delta_{jk})
		\frac{\langle M_{jn}|H_{(k-j)p}|M_{kn}\rangle}{h},\notag\\
		P_{jk}^{(p)}={}&jp\,\delta_{jk}.
		\label{eq:EMReducedMatrices}
	\end{align}
	The dressed resonance frequencies are the generalized eigenvalues
	\begin{equation}
		\det\!\left[C^{(n,p)}-f_dP^{(p)}\right]=0.
		\label{eq:EMReducedPole}
	\end{equation}
	Each commensurate family is solved independently.  The state $|M_0,0\rangle$
	sets the reference quasienergy but is not included in the excited-channel
	diagonalization, so the roots locate the solid crossing loci in
	Fig.~\ref{fig:tunneling}(c,d).  The displayed loci use $L=2$ and $L=3$,
	respectively.  The off-diagonal matrix elements in
	$C^{(n,p)}_{jk}$ are controlled by the rapidly decreasing Bessel envelope
	$J_{|j-k|p}(\phi_{\rm ac})$.
	
	For Fig.~\ref{fig:tunneling}(c), the two-channel sector
	$\{|M_{11},-1\rangle,|M_{22},-2\rangle\}$ gives a $0.8408\,{\rm GHz}$
	splitting without a fitted parameter, which agrees with the  $0.8\,{\rm GHz}$ spacing read out from Fig.~\ref{fig:tunneling}(c), whose pixel resolution is $0.1\,{\rm GHz}$. Figure~\ref{fig:tunneling}(d) uses
	$\{|M_9,-1\rangle,|M_{18},-2\rangle,|M_{27},-3\rangle\}$, which captures the splitting of the outer two states accurately. However the location of the inner state is susceptible to additional perturbation, such as the additional Sambe state $\ket{M_{36},-4}$, which shifts the location of the inner state by about $0.15\,{\rm GHz}$.

	\def\INCLUDEDSUPPLEMENT{1}
\ifdefined\INCLUDEDSUPPLEMENT
\let\finishsupplement\relax
\else
\documentclass[aps,prl,onecolumn,floatfix,nofootinbib]{revtex4-2}

\usepackage{amsmath,amssymb}
\usepackage{bm}
\usepackage[colorlinks=true,linkcolor=blue,citecolor=blue,urlcolor=blue]{hyperref}

\newcommand{\Ec}{E_{C}}
\newcommand{\EJ}{E_{J}}
\newcommand{\EL}{E_{L}}
\newcommand{\Hmov}{\mathcal{H}_{\mathrm{mov}}}
\newcommand{\xqp}{x_{\mathrm{qp}}}
\newcommand{\qpratefix}[1]{{\color{blue}#1}}
\newcommand{\newroundedit}[1]{{\color{blue}#1}}

\begin{document}

\title{Supplemental Material for ``Floquet Quasiparticle Poisoning of Frozonium''}
\author{Haoyu Guo}
\author{Debanjan Chowdhury}
\affiliation{Department of Physics, Cornell University, Ithaca, NY 14853, USA}
\date{\today}
\maketitle
\def\finishsupplement{\end{document}}
\fi

\ifdefined\INCLUDEDSUPPLEMENT
\clearpage
\onecolumngrid
\setcounter{section}{0}
\renewcommand{\thesection}{S\arabic{section}}
\renewcommand{\theHsection}{suppv2.\arabic{section}}
\setcounter{equation}{0}
\renewcommand{\theequation}{S\arabic{equation}}
\renewcommand{\theHequation}{suppv2.\arabic{equation}}
\begin{center}
{\large\bfseries Supplemental Material for ``Floquet Quasiparticle Poisoning of Frozonium''}\\[0.75em]
Haoyu Guo and Debanjan Chowdhury\\[0.35em]
Department of Physics, Cornell University, Ithaca, NY 14853, USA
\end{center}
\else
\setcounter{equation}{0}
\renewcommand{\theequation}{S\arabic{equation}}
\renewcommand{\theHequation}{suppv2.\arabic{equation}}
\fi

 This Supplemental Material contains two additional
elements: the numerical algorithm used to construct the rate maps and the
finite-width correction to the cold-edge quasiparticle approximation.

\section{Numerical construction of the rate maps}
\label{sec:suppv2-numerics}

\subsection{Floquet exact diagonalization}

We represent the circuit in a truncated oscillator basis
$\{|m\rangle_{\rm osc}:m=0,\ldots,N-1\}$ of the quadratic Hamiltonian
$4\Ec\hat n^2+\EL\hat\varphi^2/2$.  In this basis,
\begin{equation}
\hat\varphi=\varphi_{\rm zpf}(a+a^\dagger),\qquad
\hat n=i n_{\rm zpf}(a^\dagger-a),
\label{eq:suppv2OscillatorOperators}
\end{equation}
where
\begin{equation}
\varphi_{\rm zpf}=\left(\frac{2\Ec}{\EL}\right)^{1/4},\qquad
n_{\rm zpf}=\left(\frac{\EL}{32\Ec}\right)^{1/4}.
\label{eq:suppv2ZeroPointFluctuations}
\end{equation}
The nonlinear operators $\cos\hat\varphi$, $\sin\hat\varphi$,
$\cos(\hat\varphi/2)$, and $\sin(\hat\varphi/2)$ are evaluated as matrix
functions: the Hermitian matrix $\hat\varphi$ is diagonalized, the scalar
function is applied to its eigenvalues, and the result is transformed back to
the oscillator basis~\cite{Lewellen2025Frozonium}.

Let $f_C=\Ec/h$, $f_J=\EJ/h$, $f_L=\EL/h$, and
$\Phi(t)=\phi_{\rm ext}+\phi_{\rm ac}\sin(2\pi f_dt)$.  The Hamiltonian in
cycle-frequency units is assembled as
\begin{align}
\frac{\Hmov(t)}{h}={}&4f_C\hat n^2+\frac{f_L}{2}\hat\varphi^2
\notag\\[-0.25ex]
&-f_J\left[\cos\Phi(t)\cos\hat\varphi
+\sin\Phi(t)\sin\hat\varphi\right].
\label{eq:suppv2NumericalHamiltonian}
\end{align}
All matrices are stored in complex double precision.  Before propagation we
subtract the lowest eigenvalue $f_0$ of the undriven Hamiltonian in
cycle-frequency units; this changes only the common dynamical phase.

For each $(f_d,\phi_{\rm ac})$, with $T=1/f_d$, we solve
\begin{equation}
i\partial_tU(t)=2\pi\left[\frac{\Hmov(t)}{h}-f_0 I_N\right]U(t),
\qquad U(0)=I_N,
\label{eq:suppv2PropagatorEquation}
\end{equation}
by continuous-time propagation in QuSpin~\cite{PWeinberg2017,PWeinberg2019}.  The relative
and absolute solver tolerances are both $10^{-12}$.  The identity is evolved
once on the grid
\begin{equation}
t_j=\frac{jT}{N_t},\qquad j=0,\ldots,N_t-1,\qquad N_t=512,
\label{eq:suppv2TimeGrid}
\end{equation}
and also to the endpoint $T$.  Thus the same matrices $U(t_j)$ are available
for every state and are reused below.

At driven points, the one-period propagator is diagonalized according to
\begin{equation}
U(T)|v_\alpha\rangle
=e^{-i\epsilon_\alpha T/\hbar}|v_\alpha\rangle,
\label{eq:suppv2FloquetNumerics}
\end{equation}
where $\epsilon_\alpha$ is the quasienergy obtained from the principal phase
of the corresponding eigenvalue.  Numerically, $\epsilon_\alpha/h$ is stored
in GHz.  The eigenvectors are orthonormalized, and the periodic modes are
reconstructed as
\begin{equation}
|u_\alpha(t_j)\rangle
=e^{i\epsilon_\alpha t_j/\hbar}U(t_j)|v_\alpha\rangle.
\label{eq:suppv2PeriodicModeReconstruction}
\end{equation}
At $\phi_{\rm ac}=0$, the time-independent eigenenergies and eigenstates are
used directly, which fixes the static rather than the folded quasienergy
gauge.

\subsection{Floquet branch labels}

The branch labels are referenced to the leading Magnus Hamiltonian rather
than to quasienergy order.  First, $\mathcal H_M^{(0)}(\phi_{\rm ac})$ is
diagonalized in the same $N$-dimensional basis at every amplitude.  Its
eigenstates at $\phi_{\rm ac}=0$ are labeled by energy.  At each subsequent
amplitude point $i$, the energy-ordered candidates
$\{|\widetilde M_j^{(i)}\rangle\}$ are assigned to the previously continued
states by
\begin{equation}
\pi_i=\underset{\pi}{\operatorname{arg\,max}}
\sum_n\left|
\langle M_n^{(i-1)}|\widetilde M_{\pi(n)}^{(i)}\rangle
\right|^2.
\label{eq:suppv2MagnusContinuation}
\end{equation}
The maximum-weight one-to-one assignment is found with the Hungarian
algorithm.  Continuing all $N$ states gives a single amplitude-continuous
Magnus reference $\{|M_n(\phi_{\rm ac})\rangle\}$, which is independent of
$f_d$ and is shared by every frequency row.

At each point of the Floquet calculation we then form the complete overlap
matrix
\begin{equation}
W_{n\alpha}(f_d,\phi_{\rm ac})
=|\langle M_n(\phi_{\rm ac})|v_\alpha(f_d,\phi_{\rm ac})\rangle|^2
\label{eq:suppv2OverlapMatrix}
\end{equation}
and solve a second Hungarian assignment,
\begin{equation}
\Pi=\underset{\pi}{\operatorname{arg\,max}}\sum_nW_{n,\pi(n)}.
\label{eq:suppv2EDMagnusAssignment}
\end{equation}
Floquet branch $n$ is the state $\alpha=\Pi(n)$; in particular, branch 0 is
the state assigned to $M_0$.  The assignment is performed independently at
each frequency: there is no continuation along $f_d$.  Near a resonance, a
branch label therefore denotes the state selected by the global one-to-one
Magnus assignment and does not imply weak hybridization or adiabatic
quasienergy continuity.

\subsection{Micromotion matrix elements and rate reduction}

For a fixed pair of raw Floquet states, define
\begin{align}
C_{\beta\alpha}(t_j)
&=\langle\psi_\beta(t_j)|\cos(\hat\varphi/2)|\psi_\alpha(t_j)\rangle,
\notag\\
S_{\beta\alpha}(t_j)
&=\langle\psi_\beta(t_j)|\sin(\hat\varphi/2)|\psi_\alpha(t_j)\rangle,
\label{eq:suppv2BaseHalfPhaseMatrixElements}
\end{align}
where $|\psi_\alpha(t_j)\rangle=U(t_j)|v_\alpha\rangle$.  Writing
$\Phi_j=\Phi(t_j)$, the two periodic-mode matrix elements entering the
Golden-rule rates are evaluated as
\begin{subequations}
\begin{align}
m_{c,\beta\alpha}(t_j)
={}&e^{i(\epsilon_\alpha-\epsilon_\beta)t_j/\hbar}
\left[\cos\!\left(\frac{\Phi_j}{2}\right)C_{\beta\alpha}(t_j)
+\sin\!\left(\frac{\Phi_j}{2}\right)S_{\beta\alpha}(t_j)\right],
\label{eq:suppv2TimeTraceCosine}\\
m_{s,\beta\alpha}(t_j)
={}&e^{i(\epsilon_\alpha-\epsilon_\beta)t_j/\hbar}
\left[\cos\!\left(\frac{\Phi_j}{2}\right)S_{\beta\alpha}(t_j)
-\sin\!\left(\frac{\Phi_j}{2}\right)C_{\beta\alpha}(t_j)\right].
\label{eq:suppv2TimeTraceSine}
\end{align}
\label{eq:suppv2TimeTraces}
\end{subequations}
These expressions implement
$\cos[(\hat\varphi-\Phi_j)/2]$ and
$\sin[(\hat\varphi-\Phi_j)/2]$ in the same junction-phase convention as the
End Matter.  At the static point $\phi_{\rm ac}=0$, the shortcut described
above instead evaluates these matrix elements directly with
$|u_\alpha(t_j)\rangle=|v_\alpha\rangle$.

The sideband amplitudes are obtained from the endpoint-excluded discrete
Fourier transform
\begin{equation}
M_{p,\beta\alpha}^{(q)}
=\frac{1}{N_t}\sum_{j=0}^{N_t-1}
e^{i2\pi qj/N_t}m_{p,\beta\alpha}(t_j),
\qquad p=c,s,
\label{eq:suppv2DiscreteSidebandTransform}
\end{equation}
which corresponds to
$m_{p,\beta\alpha}(t)=\sum_qM_{p,\beta\alpha}^{(q)}e^{-iq\omega t}$.
For each channel, the physical frequency transferred to the quasiparticle
sector is
\begin{equation}
f_{\alpha\beta}^{(q)}
=\frac{\epsilon_\alpha-\epsilon_\beta}{h}+qf_d.
\label{eq:suppv2NumericalTransferFrequency}
\end{equation}
The squared sine and cosine amplitudes and this physical transfer frequency
are inserted into the pair-generation or tunneling formula given in the End
Matter.  Pair generation uses the two elliptic BCS form factors
$S_{\rm gen}^{c,s}$ and vanishes for
$f_{\alpha\beta}^{(q)}<2f_\Delta$.  Pre-existing-quasiparticle tunneling uses
both finite-transfer cold-edge factors $S_{\rm tun}^{s,c}$ and only channels
with $f_{\alpha\beta}^{(q)}>0$.  The tunneling map is reported per unit
$\xqp$.

For each of the $N$ initial states, pair generation and tunneling are summed
over the sideband sets $\mathcal Q_{\rm pair}$ and
$\mathcal Q_{\rm tunnel}$, respectively, as listed in
Table~\ref{tab:suppv2NumericalParameters}.  Both calculations sum over all
$N$ final Floquet states.  The state-changing map associated with branch $n$
is
\begin{equation}
\Gamma_n^X(f_d,\phi_{\rm ac})
=\sum_{\beta\ne\Pi(n)}\sum_{q\in\mathcal Q_X}
\Gamma_{\Pi(n)\to\beta}^{X,(q)},
\qquad X\in\{{\rm pair},{\rm tunnel}\}.
\label{eq:suppv2StateChangingReduction}
\end{equation}
Contributions with identical raw initial and final states are accumulated separately and excluded from the state-changing rate in
Eq.~\eqref{eq:suppv2StateChangingReduction}.

The amplitude grid is $\phi_{\rm ac}=0,0.0125,\ldots,3$, supplemented by the
exact first zero $\phi_{\rm ac}^\star=2.4048255577$ of $J_0$, for 242
amplitudes in total.  The calculation contains 516 frequency rows, including
the complete $0.1\,{\rm GHz}$ grid from $5$ to $50\,{\rm GHz}$ and additional
rows up to $92\,{\rm GHz}$.  Every map pixel is obtained directly from
Eq.~\eqref{eq:suppv2StateChangingReduction}; the rate calculation does not
interpolate between parameter points.

For the dashed static-Magnus curves in Fig.~2(b), the same matrix-element,
sideband, form-factor, and final-state sums are repeated with
$|u_\alpha(t)\rangle$ replaced by $|M_\alpha\rangle$.  The explicit time
dependence of the half-phase operators is retained, while circuit-state
micromotion and resonant Floquet hybridization are omitted.  Thus the dashed
curves differ from the ED result only through the circuit states used in the
matrix elements and transition frequencies.

\begin{table}[htb]
\caption{Numerical parameters used to construct the ED rate maps.}
\label{tab:suppv2NumericalParameters}
\begin{ruledtabular}
\begin{tabular}{p{0.30\textwidth}c p{0.42\textwidth}}
Quantity & Symbol or setting & Value\\
\colrule
{Josephson energy} & {$\EJ/h$} &
{$4.88\,\mathrm{GHz}$}\\
{Charging energy} & {$\Ec/h$} &
{$1.09\,\mathrm{GHz}$}\\
{Inductive energy} & {$\EL/h$} &
{$0.56\,\mathrm{GHz}$}\\
Hilbert-space dimension & $N$ & $200$\\
Time samples per period & $N_t$ & $512$, endpoint excluded\\
Relative solver tolerance & \texttt{rtol} & $10^{-12}$\\
Absolute solver tolerance & \texttt{atol} & $10^{-12}$\\
Pair-generation sidebands & $\mathcal Q_{\rm pair}$ & $q=0,\ldots,30$\\
QP-tunneling sidebands & $\mathcal Q_{\rm tunnel}$ & $q=-30,\ldots,30$\\
Amplitude grid & $\phi_{\rm ac}$ &
$0:0.0125:3$, plus $\phi_{\rm ac}^{\star}=2.4048255577$; $242$ points\\
Frequency grid & $f_d$ &
$516$ rows, including the $0.1\,{\rm GHz}$ grid from $5$ to $50\,{\rm GHz}$
and additional rows to $92\,{\rm GHz}$\\
\end{tabular}
\end{ruledtabular}
\end{table}

\section{Finite-width correction to quasiparticle tunneling}
\label{sec:suppv2-finite-width}

The End Matter evaluates tunneling of pre-existing quasiparticles in the
cold-edge limit, where their occupied energies lie sufficiently close to the
gap that the initial energy can be set to $E=\Delta$ in the spectral factors.
Here we retain the energy width of a general dilute, stationary occupation
$f_E(E)$.  Set
\begin{equation}
x=\frac{E}{\Delta}\ge1,\qquad
w=\frac{hf}{\Delta}=\frac{f}{f_\Delta},\qquad
\xqp=2\int_1^\infty dx\,
\frac{x}{\sqrt{x^2-1}}f_E(\Delta x).
\label{eq:suppv2Definitions}
\end{equation}
Carrying out the BCS momentum sums without expanding in $w$ gives
\begin{subequations}
\begin{align}
S_{\rm tun}^{s}(f)
={}&\frac{16\EJ}{\pi{\hbar}}\int_1^\infty dx\,
f_E(\Delta x)\{1-f_E[\Delta(x+w)]\}
\notag\\[-0.25ex]
&\times\frac{x(x+w)+1}
{\sqrt{x^2-1}\sqrt{(x+w)^2-1}},
\label{eq:suppv2SqpSExact}\\
S_{\rm tun}^{c}(f)
={}&\frac{16\EJ}{\pi{\hbar}}\int_1^\infty dx\,
f_E(\Delta x)\{1-f_E[\Delta(x+w)]\}
\notag\\[-0.25ex]
&\times\frac{x(x+w)-1}
{\sqrt{x^2-1}\sqrt{(x+w)^2-1}}.
\label{eq:suppv2SqpCExact}
\end{align}
\label{eq:suppv2SqpExact}
\end{subequations}
The upper and lower coherence factors follow from the particle-hole sums
$|B^+|^2\to EE'+\Delta^2$ and
$|B^-|^2\to EE'-\Delta^2$, with $E'=E+hf$.  Concentrating $f_E(E)$ at the
gap edge reduces Eqs.~\eqref{eq:suppv2SqpExact} to the cold-edge form factors
quoted in the End Matter.

To quantify the correction, we take a dilute Boltzmann occupation as a
quasiequilibrium benchmark~\cite{Connolly2024Quasiequilibrium}; nonequilibrium
quasiparticle distributions need not in general be described by a single
effective temperature~\cite{Serniak2018HotQPs}:
\begin{equation}
f_E(E)=\exp\!\left[-\frac{E}{k_{\rm B}T_{\rm eff}}\right],\qquad
\xqp(T_{\rm eff})=2K_1\!\left(\frac{\Delta}{k_{\rm B}T_{\rm eff}}\right),
\label{eq:suppv2BoltzmannNormalization}
\end{equation}
where $K_1$ is the modified Bessel function.  At fixed $\xqp$, the exact
spectra are
\begin{equation}
S_{\rm tun}^{s,c,{\rm B}}(f)
=\xqp\frac{8\EJ}{\pi{\hbar}}
\frac{1}{K_1\!\left(\Delta/k_{\rm B}T_{\rm eff}\right)}
\int_1^\infty dx\,e^{-x\Delta/k_{\rm B}T_{\rm eff}}
\frac{x(x+w)\mathbin{\pm}1}
{\sqrt{x^2-1}\sqrt{(x+w)^2-1}},
\label{eq:suppv2SqpFiniteWidth}
\end{equation}
with the upper sign for the sine channel and the lower sign for the cosine
channel.  The Pauli-blocking factor has been set to one consistently with the
dilute approximation.  For $k_{\rm B}T_{\rm eff}\ll\Delta$,
\begin{equation}
\xqp(T_{\rm eff})
\simeq\sqrt{\frac{2\pi k_{\rm B}T_{\rm eff}}{\Delta}}
\exp\!\left[-\frac{\Delta}{k_{\rm B}T_{\rm eff}}\right].
\label{eq:suppv2BoltzmannLowTemperature}
\end{equation}

For $f_\Delta=45\,{\rm GHz}$, the resulting effective temperatures and the
distribution of the fractional reduction
$1-\Gamma_0^{{\rm tun},{\rm B}}/\Gamma_0^{{\rm tun},{\rm ce}}$ over the
$5$--$50\,{\rm GHz}$ branch-0 scan are
\begin{equation}
\begin{array}{c|c|c|c|c}
\xqp
&T_{\rm eff}\ ({\rm mK})
&\text{Median}
&\text{90th percentile}
&\text{Largest}\\ \hline
10^{-6}&160&2.95\%&22.4\%&77.8\%\\
10^{-7}&138&2.58\%&21.1\%&77.0\%\\
10^{-8}&121&2.29\%&20.0\%&76.2\%
\end{array}.
\label{eq:suppv2ThermalRateSensitivity}
\end{equation}
The correction to the total branch-0 decay rate is smaller when pair
generation dominates.  {At the sampled point nearest the operating region,
$f_d=21.7\,{\rm GHz}$, finite width reduces the total rate by $1.62\%$.}  {At
$f_d=11.8\,{\rm GHz}$, where pre-existing-quasiparticle tunneling dominates,
it reduces the total rate by $66.4\%$ and increases the branch-0 lifetime from
$2.03$ to $6.06\,\mu{\rm s}$.}  Thus the cold-edge approximation
overestimates the quasiparticle-induced decay rate and correspondingly
underestimates the lifetime.

\ifdefined\INCLUDEDSUPPLEMENT\else
\bibliography{references}
\fi

\finishsupplement

\end{document}